\documentclass[aps,pre,groupedaddress]{revtex4-2}
\usepackage{natbib}
\usepackage{amsmath}
\usepackage{epsfig}
\usepackage{color}

\usepackage{graphicx}
\usepackage{amsfonts}
\usepackage{dcolumn}
\usepackage{bm}
\usepackage{color}
\usepackage{hyperref}

\begin{document}
\bibliographystyle{unsrt}

\title{Semi-Markov Processes in Open Quantum Systems. II. Counting Statistics with Resetting} 
\author{Fei Liu}
\email[Email address: ]{feiliu@buaa.edu.cn}
\affiliation{School of Physics, Beihang University, Beijing 100083, China}

\date{\today}

\begin{abstract}
A semi-Markov process method for 
obtaining general counting statistics 
for open quantum systems is extended to the scenario of resetting. The simultaneous presence of random resets and wave function collapses means that the quantum jump trajectories are no longer semi-Markov.
However, focusing on trajectories and using simple probability formulas, general counting statistics can still be constructed from reset-free statistics. An exact tilted matrix equation is also obtained. The inputs of these methods are the survival distributions and waiting-time density distributions instead of quantum operators. In addition, a continuous-time cloning algorithm is introduced to simulate the large-deviation properties of open quantum systems. Several quantum optics systems are used to demonstrate these results.  
\end{abstract}

\maketitle

\section{Introduction} 
In a previous paper~\cite{Liu2022}, we explicitly constructed semi-Markov processes (sMPs) embedded in quantum jump trajectories of open quantum systems and clarified their connections with the Markov quantum master equation (MQME)~\cite{Breuer2002,Rivas2012,Alicki2010}.  
The unique advantages of the sMP method are in analyses and computations of the general counting statistics of open quantum systems~\cite{Levitov1996,Bagrets2003,Esposito2009,Rudge2019}, e.g., the large deviation properties of these systems. On the one hand, unlike the tilted quantum master equation (TQME)~\cite{Mollow1975,Carmichael1989,Zoller1987,Zheng2003,Garrahan2010,Esposito2009,Liu2016a}, which is now dominant in the literature, the sMP method is a theory concerning classic probability; all quantum characteristics are indirectly indicated through the classical waiting time distributions. 
On the other hand, the sMP method can handle the statistics of general time-extensive quantities related to the occurrence frequencies of adjacent collapses in quantum jump trajectories. In contrast, the TQME is restricted in the time-extensive quantities of the frequencies of single collapses. Notably, the counting statistics of sMPs were established over a decade~(\cite{Andrieux2008,Esposito2008}). However, to the best of our knowledge, their significance in open quantum systems has not been appreciated until the present work.
  
In this paper, we aim to deepen the sMP method by investigating the counting statistics in more complex open quantum systems in the presence of stochastic resetting. In the past decade, stochastic systems with various resetting protocols have attracted much theoretical interest in the community of nonequilibrium physics ~\cite{Evans2011,Gupta2014,Meylahn2015,Eule2016,Chechkin2018,Gupta2020,Pal2021,Gupta2022}. Although most of this work has involved classical systems, some attention has also been devoted to quantum systems~\cite{Hartmann2006,Linden2010,Rose2018,Mukherjee2018,Tavakoli2020,Carollo2019,Perfetto2021,Perfetto2022,Yin2023}, e.g., constructing autonomous entanglement engines utilizing resetting~\cite{Hartmann2006,Tavakoli2020}, designing nonequilibrium stationary states of quantum many-body systems through resetting~\cite{Perfetto2021}, and achieving speedup of quantum hitting times by resetting~\cite{Yin2023}. Very recently, Perfetto et al.~\cite{Perfetto2022} studied the thermodynamics of quantum jump trajectories subject to non-Poissonian resetting. They found that the large deviation properties of the counting statistics can be calculated exactly by relating the moment-generating function (MGF) in the presence of resetting to that of a reset-free system. To achieve this result, they combined techniques used on the TQME with the renewal structure of the resetting dynamics. 

The presence of stochastic resetting in open quantum systems raises a fundamental challenge to the sMP method. In the reset-free case, according to the theory of quantum jump trajectories~\cite{Mollow1975,Srinivas1981,Zoller1987,Dalibard1992,Carmichael1993,Plenio1998,Breuer2002,Gardiner2004,Wiseman2010}, the wave function of the quantum system deterministically evolves in a nonunitary way and is randomly interrupted by collapses. Because the time distribution of pairs of adjacent collapses is non-Poissonian, called memory in this paper, and is independent of the previous history of the wave function, if we are only concerned about collapses of wave functions, including the collapsed quantum states and times, these random events constitute a sMP~\cite{Ross1995,Qian2006}. In general, the resetting process is also an sMP. Hence, when these two stochastic processes occur simultaneously, the composite process is usually no longer semi-Markovian.  

This work overcomes the aforementioned challenge, and it makes three main contributions. First, we construct a set of probability formulas that can precisely describe the composite stochastic process. The key idea is that resetting does not alter the underlying quantum dynamics; the notion of quantum jump trajectories is still valid. Second, we extend the previous results of Perfetto et al.~\cite{Perfetto2022} to general counting statistics. Because our theory is fully based on sMPs, previously unknown formulas are discovered. For simplicity of description, the counting statistics mentioned in the remainder of this paper always refer to the general statistics unless otherwise indicated. Finally, we introduce a continuous-time cloning algorithm (CTCA) to simulate the large deviation statistics of general time-extensive quantities in open quantum systems with resetting. The algorithm originally aimed to compute the scaled cumulant generating functions (SCGFs) of non-Markov classical jump processes~\cite{Cavallaro2016,Lecomte2007,Giardina2006}. Because the set of probability formulas for the composite stochastic process is obtained, the applications of this method in open quantum systems are natural, and its realization is also simple.  

This paper is organized as follows. In Sec.~(\ref{section2}), we briefly summarize the sMP method for determining the counting statistics of open quantum systems and extend it to a situation with arbitrary initial states. In Sec.~(\ref{section3}), counting statistics with memoryless resetting are studied. We will see that the method is still available if the set of collapsed states is expanded to include the reset state. Section~(\ref{section4}) discusses a more complex case with memory resetting. Although quantum jump trajectories in this situation are no longer sMPs, by considering trajectories and using probability formula, the counting statistics can still be constructed by reset-free statistics. In particular, an exact tilted matrix equation is obtained. In Sec.~(\ref{section5}), we introduce a continuous-time cloning algorithm to simulate the large-deviation statistics. In Sec.~(\ref{section6}), several quantum optics systems are used to demonstrate our results. Section~(\ref{section7}) concludes the paper.
  
\section{SMP method for counting statistics}
\label{section2}
Let $\rho(t)$ be the reduced density matrix of an open quantum system. Under appropriate conditions, the dynamics of the system is described by the  MQME~\cite{Davies1974,Lindblad1976,Gorini1976} 
\begin{eqnarray}
	\label{MQME}
	\partial_t \rho(t)=-{\rm i}[H,\rho(t)]+\sum_{\alpha=1}^M r_\alpha\left( A_\alpha\rho(t)A^\dag_\alpha -\frac{1}{2}\left\{A^\dag_\alpha A_\alpha,\rho(t)\right\}\right)\equiv {\cal L}[\rho(t)],
\end{eqnarray}
where the Planck constant $\hbar$ is set to 1, $H$ denotes the Hamiltonian of the quantum system, $A_\alpha$ is the Lindblad operator, and the nonnegative coefficients $r_\alpha$ and $\alpha=1,\cdots,M$ represent certain correlation characteristics of the environment surrounding the quantum system. 

The MQME~(\ref{MQME}) can be unraveled into quantum jump trajectories~\cite{Mollow1975,Srinivas1981,Zoller1987,Dalibard1992,Carmichael1993,Plenio1998,Breuer2002,Gardiner2004,Wiseman2010}. These trajectories, which concern the evolutions of the wave functions of the single quantum systems, are composed of deterministic pieces and random collapses of the wave functions. The former are the solutions of nonlinear Schr$\ddot{o}$dinger equations. The latter indicate that the systems have collapsed to fixed states $\phi_\alpha$, $\alpha=1,\cdots, M$, which are called the collapsed states in this paper. If we focus on these states and use random time intervals $\tau$ to replace the deterministic pieces between successive collapses, the quantum jump trajectories can be seen as the realizations of a sMP~\cite{Liu2022}. The ingredients of the sMP include the waiting time densities (WTDs) and survival distributions (SDs)~\cite{Qian2006,Ross1995}, which are
\begin{eqnarray}
\label{waitingtimedensity}
	p^{0}_{\alpha|\beta}(\tau)&=&r_\beta\parallel A_\beta e^{-{\rm i}\tau \hat H}\phi_\alpha \parallel^2,\\ 
	S^{0}_\alpha(\tau)&=&\parallel e^{-{\rm i}\tau \hat H}\phi_\alpha\parallel^2,
	\label{survivaldistribution}
\end{eqnarray}
respectively~\cite{Breuer2002}. Here, the non-Hermitian Hamiltonian is  
\begin{eqnarray}
\hat H=H-\frac{\rm i}{2}\sum_{\alpha=1}^M r_\alpha A_\alpha^\dag A_\alpha.
\end{eqnarray}
Equation~(\ref{waitingtimedensity}) is the probability density of the wave function starting from collapsed state $\phi_\alpha$, continuously evolving, and collapsing in state $\phi_\beta$ until time $\tau$. Equation~(\ref{survivaldistribution}) is the probability of the wave function successively evolving until time $\tau$ without collapse. In this paper, we always denote quantities defined or solved in the absence of resetting with a superscript or subscript $0$, unless otherwise stated. It is useful to introduce the hazard functions of the sMP:
\begin{eqnarray}
	k^0_{
		\alpha|\beta}(\tau)&=&\frac{p^0_{\alpha|\beta}(\tau)}{S_\alpha^0(\tau)}.
\end{eqnarray}
Obviously, they are the conditional probability densities at which the system collapses in $\phi_\beta$ at time $\tau$ within a unit time interval under the condition that the system continuously evolves from state $\phi_\alpha$ until time $\tau$ without collapsing.  

A major application of the sMP perspective to open quantum systems is in counting statistics~\cite{Levitov1996,Bagrets2003,Esposito2009,Rudge2019}. These statistics concern time-extensive quantities     
\begin{eqnarray}
	\label{generalcountingquantity}
	C[\vec X]=\sum_{i=1}^N \omega_{\alpha_{i-1}\alpha_i}.
\end{eqnarray}
Here, we denote the quantum jump trajectory with $N$ collapses as 
\begin{eqnarray}
\vec X=(\phi_{\alpha_1},\phi_{\alpha_2},\cdots,\phi_{\alpha_N}),
\end{eqnarray}
where $\phi_{\alpha_i}$ represents the collapsed state at time $t_i$, $i=1,\cdots,N$, and $\omega_{\alpha_{i-1}\alpha_i}$ is a weight specified by the collapsed states at adjacent times $t_{i-1}$ and $t_i$, that is, the quantum states at the beginning and end times of a deterministic stage. The properties of the random variable~(\ref{generalcountingquantity}) are characterized by the MGF~\cite{vanKampen2007} 
\begin{eqnarray}
	\label{momentgeneratingfunction}
	M_0(\lambda,t)=\sum_{\vec X} {\cal P}[\vec X]e^{-\lambda C[\vec X]},
\end{eqnarray}
where ${\cal P}[\vec X]$ represents the probability density of a quantum jump trajectory $\vec X$. We set all the trajectories to start from a certain collapsed state. Note that the summation in Eq.~(\ref{momentgeneratingfunction}) over quantum jump trajectories is only a shorthand notation, and its exact meanings include summing over all possible collapsed states at every time and performing time-ordered integrals at different times; see also Eq.~(\ref{timeintegrals}) below.

The MGF~(\ref{momentgeneratingfunction}) is obtained by first solving a tilted matrix equation in the complex frequency domain~\cite{Liu2022}:
\begin{eqnarray}
	\label{MEofcountingstatisticsmatrixform}
	{\bf G}_0(v)\hat {\bf P}_0(v)={\bf 1}_\gamma,
\end{eqnarray}
where the $1\times M$ vector $ \hat {\bf P}_0^T=(\hat P_1,\cdots,\hat P_M)$ with the uppercase $T$ denoting the transpose, ${\bf 1}_\gamma^T=(\delta_{1\gamma},\cdots,\delta_{M\gamma})$ and $\delta_{\alpha\gamma}$ is the Kronecker symbol. Here, the initial collapsed state is set to $\phi_\gamma$. Throughout this paper, we use a circumflex placed over a symbol to denote its Laplace transformation. The diagonal and nondiagonal elements of the matrix ${\bf G}_0$ are 
\begin{eqnarray}
	{[{\bf G}_0]}_{\alpha\alpha}&=&\frac{1-\hat{p}^0_{\alpha|\alpha}(v)e^{-\lambda\omega_{\alpha\alpha}} }{\hat S^0_\alpha(v)},\\
 	{[{\bf G}_0]}_{\alpha\beta}&=&-\frac{\hat{p}^0_{\beta|\alpha}(v)}{\hat{S}^0_\beta(v)}e^{-\lambda\omega_{\beta\alpha}}\hspace{0.5cm} (\alpha\neq\beta),
\end{eqnarray}
respectively. Then, the Laplace transform of the MGF is 
\begin{eqnarray}
	\label{MGFresettingfreenonresettinginitialstate}
{\hat M}_0(\lambda,v)
={\bf 1}^T\hat{\bf P}_0(v)={\bf 1}^T{\bf G}^{-1}_0(v){\bf 1}_\gamma,
\end{eqnarray}
where ${\bf 1}^T$$=$$(1,\cdots,1)$ is a $1\times M$ vector. The last step is to take the inverse Laplace transform of Eq.~(\ref{MGFresettingfreenonresettinginitialstate}) over the complex frequency $v$ to obtain the MGF in the time domain. 
Equation~(\ref{MGFresettingfreenonresettinginitialstate}) is itself useful since the SCGF of the large deviation of current $j=C[\vec X]/t$ over a long time limit \cite{Touchette2008},
\begin{eqnarray}
	\varphi(\lambda)&=&\lim_{t\rightarrow\infty} \frac{1}{t}\ln M_0(\lambda,t),
\end{eqnarray} 
can be obtained by finding its pole with the largest real value~\cite{Andrieux2008}. According to Eq.~(\ref{MEofcountingstatisticsmatrixform}), it is also equal to the largest real root of the vanishing determinant of the tilted matrix ${\bf G}_0(v)$.

\subsection{Arbitrary initial states}
\label{subsection1}
We extend the previous results to a situation in which the quantum jump trajectories start with a quantum state that does not belong to the set of collapsed states, e.g., $|A\rangle$~\footnote{This limitation is not  essential and will be removed in the following sections}. For an ``autonomous" quantum system, if it starts with such a state, after the first collapse, all subsequent collapse states are in the set of 
collapsed states; that is, the system will never collapse to $|A\rangle$ again. Because the quantum jump trajectories are still a sMP, we expand
the previous vector and matrix to $1\times(M+1)$ $ \hat {\bf P}_{0A}^T=(\hat P_A,\hat P_1,\cdots,\hat P_M)$ and $(M+1)\times(M+1)$ ${\bf G}_{0A}(v)$, respectively. Note that we use the subscript $A$ to indicate the arbitrary initial state. The elements of the matrix are  
\begin{eqnarray}	
	&&[{\bf G}_{0A}]_{00}=\frac{1}{\hat S_A^0(v)}, \hspace{1.8cm}
	[{\bf G}_{0A}]_{0\beta}= 0, \label{G0arbitraryinitialstate1}\\
	&&[{\bf G}_{0A}]_{\alpha 0}=-\frac{\hat p^0_{A|\alpha}(v)}{\hat S_A^0(v)}e^{-\lambda \omega_{A\alpha}},\hspace{0.2cm}[{\bf G}_{0A}]_{\alpha \beta}=[{\bf G}_0]_{\alpha \beta}(v)
	\label{G0arbitraryinitialstate2}.  
\end{eqnarray} 
In these equations, $p^0_{A|\alpha}$ and $S_A^0$ possess analogous formulas to Eqs.~(\ref{waitingtimedensity}) and~(\ref{survivaldistribution}) except that the subscripts $\alpha$ and $\beta$ are replaced by $A$ and $\alpha$, respectively. An analogous tilted matrix equation about ${\bf G}_{0A}$ and $\hat {\bf P}_{0A}^T$ such as   Eq.~(\ref{MEofcountingstatisticsmatrixform}) is valid as well. Therefore, the Laplace transform of the MGF with the special initial state is  
\begin{eqnarray}
	\label{MGFresettingfreearbtraryinitialquantumstate}
	\hat{M}_{0A}(\lambda,v)={\bf 1}^T\hat {\bf P}_{0A}(v)={\bf 1}^T{\bf G}_{0A}^{-1}(v) {\bf 1}_A. 
\end{eqnarray}
Here, both ${\bf 1}^T$ and ${\bf 1}^T_A=(1,0,\cdots,0)$ are $1\times (M+1)$ vectors. Because the determinants of the tilted matrices ${\bf G}_0$ and ${\bf G}_{0A}$ are the same, an arbitrary initial state does not alter the large deviation properties of open quantum systems.  
 
\section{Counting statistics with memoryless resetting}
\label{section3}
\begin{figure}
	\includegraphics[width=1\columnwidth]{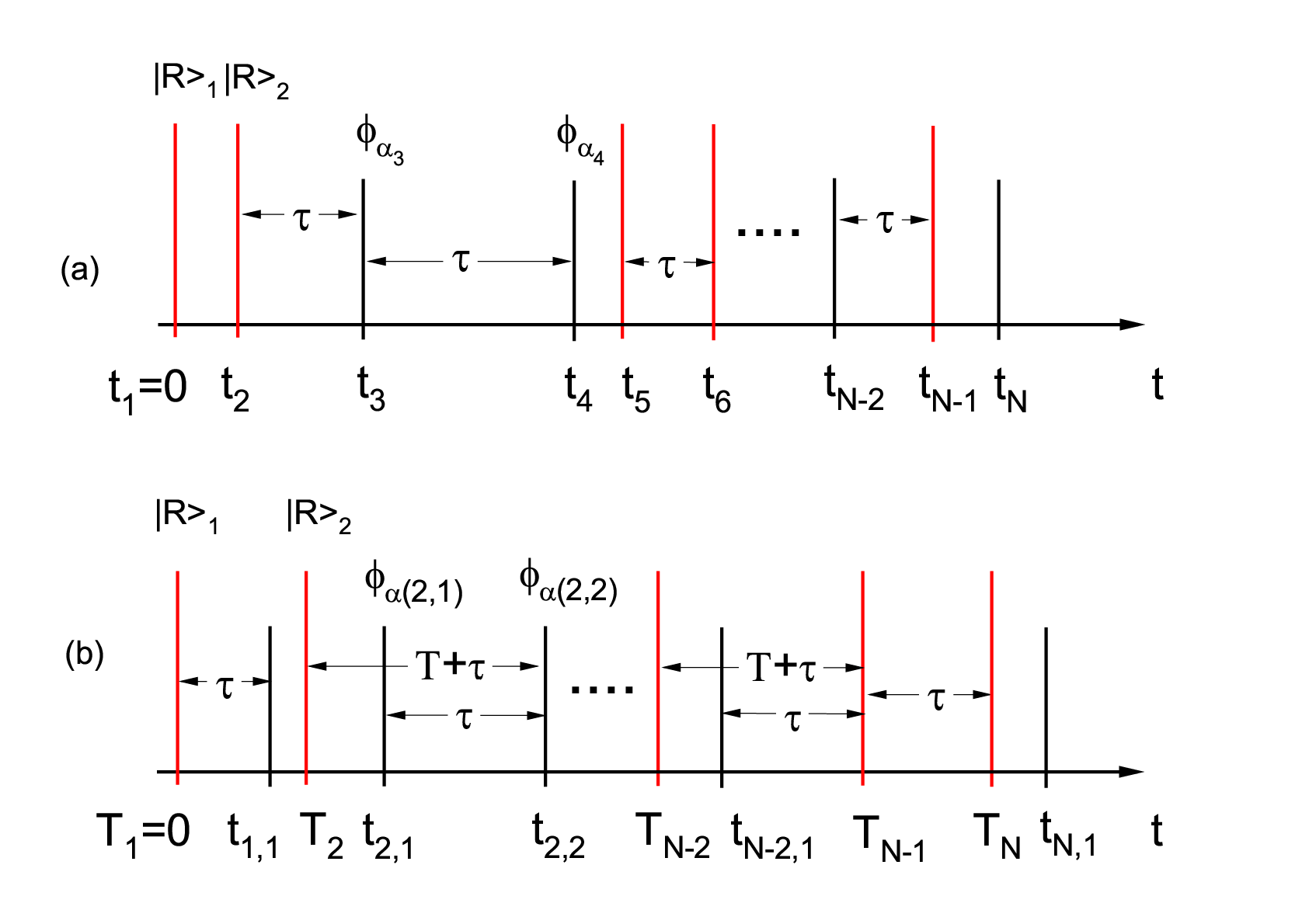}
	\caption{Schematic diagrams of two quantum jump trajectories of a quantum system with memoryless resetting (a) and memory resetting (b). The long red vertical lines represent the times at which resetting occurs, while the short black vertical lines represent the times at which collapses of wave functions occur. In Panel (a), the times are indicated by $t_i$. The collapsed states and the reset state at these times are denoted by $\phi_{\alpha_i}$ and $|R\rangle_i$, respectively, $i=1,2,\cdots,N$. There are four types of combinations of different beginning and end states. We mark them in the panel by horizontal short lines with double arrows. Their time intervals are uniformly labeled as $\tau$. In Panel (b), we denote the resetting times as $T_I$, $I=1,\cdots,N$, while the times at which the quantum system collapses are labeled as $t_{I,i}$. The subscripts indicate the $i$-th collapse following the $I$-th resetting, where $i=1,2,\cdots$. Accordingly, the collapsed states are denoted as $\phi_{\alpha(I,i)}$. There are also four types of combinations of different beginning and end states. We still denote them with horizontal lines with double arrows. Because the time interval $\tau$ is inadequate to characterize the memory effects on the two combinations, we additionally define $T$, which is the time interval from the first collapsed state to the last resetting.  }
	\label{fig1}
\end{figure}
Let us start with the resetting case in which resetting occurs at a constant rate $K$ and is independent of previous quantum states. This is a special case of more general memory resetting. However, we will prove that this case can be studied by simply modifying the reset-free results. Throughout this paper, we only consider one type of resetting, and quantum systems are always reset to $|\phi_R\rangle$. This resetting quantum state may be one of the collapsed states. In addition, we initialize all quantum trajectories to the reset state. 

Because resetting is memoryless, the quantum jump trajectories in the case of a reset state are still sMPs. This point is shown schematically in Fig.~(\ref{fig1})(a). At time $t_2$, resetting occurs. Then, the evolution of the quantum system that started from the collapsed state $\phi_{\alpha_3}$ at time $t_3$ is not affected by the previous wave function history, including the time interval $t_3-t_2$ and the $t_2$ value. Therefore, we naturally think of the quantum jump trajectories as having a set of extended collapsed states; that is, the resetting quantum state $|R\rangle$ is included. On the other hand, we emphasize that resetting indeed leads to new SDs and WTDs and modifies the existing SDs and WTDs:    
\begin{eqnarray}
	\label{survivaldistributionmemorylessresetting}
	&&S_\alpha(\tau)=e^{-\int_0^\tau \sum_{\beta=1}^M k^0_{\alpha|\beta}ds-K\tau }=S_\alpha^0(\tau)e^{-K\tau},  \\
    &&S_R(\tau)=e^{-\int_0^\tau \sum_{\beta=1}^M k_{R|\beta}^0ds-K\tau}=S_R^0(\tau)e^{-K\tau},
  \label{survivaldistributionR0}  
\end{eqnarray} 
and 
\begin{eqnarray}
	&&p_{\alpha|\beta}(\tau)=k^0_{\alpha|\beta}(\tau)S_\alpha(\tau)=p^0_{\alpha|\beta}(\tau)e^{-K\tau}, \\
	&&p_{\alpha|R}(\tau)=KS_\alpha(\tau)=S^0_{\alpha}(\tau)Ke^{-K\tau},\\
	&&p_{R|\beta}(\tau)=k^0_{R|\beta}(\tau) S_R(\tau)=p^0_{R|\beta}(\tau)e^{-K\tau},\label{waitingtimedispRbeta}\\
	&&p_{R|R}(\tau)=K S_R(\tau)=S^0_R(\tau)Ke^{-K\tau}.\label{waitingtimedispRR}
\end{eqnarray} 

With Eqs.~(\ref{survivaldistributionmemorylessresetting})-(\ref{waitingtimedispRR}), the MGF in the presence of memoryless resetting is calculated in the same way as in the previous reset-free case. A difference is that in the current case, the tilted matrix ${\bf G}$ is $(M+1)\times(M+1$), the elements of which are  
\begin{eqnarray}
	&&[{\bf G}]_{00}=\frac{1}{\hat S^0_R(v+K)} -K, \hspace{1.cm}
	[{\bf G}]_{0\beta}= -K,\nonumber \\
	&&[{\bf G}]_{\alpha 0} = -\frac{\hat p^0_{R|\alpha}(v+K)}{\hat S^0_R(v+K)}e^{-\lambda\omega_{R\alpha}},\hspace{0.2cm}[{\bf G}]_{\alpha \beta}=[{\bf G}_0]_{\alpha \beta}(v+K).\nonumber 
\end{eqnarray} 
Because we are not interested in the number of resets, we set the weights $\omega_{RR}$ and $\omega_{\alpha R}$ to zero. Obviously, the tilted matrices defined thus far have a simple relationship: 
\begin{eqnarray}
	\label{tiltedmatrixmemorylessresetting}
	{\bf G}(v)= {\bf G}_{0R}(v+K) -K{\bf\Pi },   
\end{eqnarray}
where the matrix elements of ${\bf G}_{0R}$ are given in Eqs.~(\ref{G0arbitraryinitialstate1}) and~(\ref{G0arbitraryinitialstate2})) with the subscript $A$ replaced by $R$, and ${\bf \Pi}=({\bf 1}_R,\cdots,{\bf 1}_R)$ is a $(M+1)\times(M+1)$ square matrix.  

Analogous to the reset-free case, the Laplace transform of the MGF in the presence of memoryless resetting is $\hat M (\lambda,v)={\bf 1}^T\hat{\bf P}(v)$, where the vector $ \hat {\bf P}^T=(\hat P_R,\hat P_1,\cdots,\hat P_M)$ satisfies the tilted matrix equation   
\begin{eqnarray}
	\label{tiltedmatrixequationwithmemorylessrestting}
	{\bf G}(v)\hat{\bf P}(v)&=&{\bf 1}_R. 
\end{eqnarray}
Substituting Eq.~(\ref{tiltedmatrixmemorylessresetting}) into Eq.~(\ref{tiltedmatrixequationwithmemorylessrestting}), multiplying both sides of the equation by ${\bf 1}^T{\bf G}_{0R}^{-1}(v+K)$ from the left-hand side, and using Eq.~(\ref{MGFresettingfreearbtraryinitialquantumstate}), we arrive at 
\begin{eqnarray}
	\label{MGFmemorylessrestting}
	\hat M (\lambda,v)
	=\frac{\hat M_{0R}(\lambda,v+K)}{1-K \hat M_{0R}(\lambda,v+K)}.
\end{eqnarray}
This result indicates that the MGF in the presence of memoryless resetting with reset state $|R\rangle$ is related to the MGF in the absence of resetting but with the special initial state $|R\rangle$.
Equation~(\ref{MGFmemorylessrestting}) also implies that the SCFG of the former case can be obtained by finding the largest real root of the following algebraic equation with the parameter $v$:  
\begin{eqnarray}
	\label{SCFGmemorylessresetting}
K\hat  M_{0R}(\lambda,v+K)-1=0.
\end{eqnarray}

\section{Counting statistics with memory resetting}
\label{section4}
We move to a more complex case with memory resetting. Unlike the previous case with memoryless resetting, a reset affects the evolution of wave functions even if the quantum system restarts from an independent collapsed state after resetting. We illustrate this point in Fig.~(\ref{fig1})(b). Assume that a reset and a subsequent collapse to the state $\phi_{\alpha(2,1)}$ occur at $T_2$ and $t_{2,1}$, respectively. Because of memory, the possibility of the continuous evolution of the quantum system that restarts from the collapsed state depends on the time interval $t_{2,1}-T_2$. 

This characteristic can be indicated by precise formulas. First, we let the hazard function of memory resetting be ${\cal K}(\tau)$. The corresponding WTD ${\cal Q}(\tau)$ of the memory resetting process is  
\begin{eqnarray}
{\cal 	Q}(\tau)={\cal K}(\tau){\cal S}(\tau),
\end{eqnarray}
and the SD is ${\cal S}(\tau)=\exp[{-\int_0^\tau {\cal K}(s)ds}]$. 
Then, we can write the WTDs and SDs of the quantum jump trajectories in the presence of memory resetting in an analogous way to Eqs.~(\ref{survivaldistributionmemorylessresetting})-(\ref{waitingtimedispRR}): 
\begin{eqnarray}
	\label{SDalphamemoryresetting}
	&&S_\alpha(\tau,T)=e^{-\int_0^\tau \sum_{\beta=1}^M k^0_{\alpha|\beta}ds-\int_{T}^{T+\tau} {\cal K}(s)ds}=S_\alpha^0(\tau)\frac{{\cal S}(T+\tau)}{{\cal S}(T)},\nonumber \\
	&&S_R(\tau)=S_R^0(\tau)e^{-\int_0^\tau {\cal K}(s)ds}=S_R^0(\tau){\cal S}(\tau),
\end{eqnarray}
and the WTDs are
\begin{eqnarray}
	\label{WTDalphabetamemoryresetting}
	&&p_{\alpha|\beta}(\tau,T)=k^0_{\alpha|\beta}(\tau)S_\alpha(\tau,T)=p_{\alpha|\beta}^0(\tau)\frac{{\cal S}(T+\tau)}{{\cal S}(T)},\\
	&&p_{\alpha|R}(\tau,T)={\cal K}(\tau+T)S_\alpha(\tau,T)=S_\alpha^0(\tau) \frac{{\cal Q}(T+\tau)}{{\cal S}(T)},\\
	&&p_{R|\beta}(\tau)=k_{R|\beta}^0(\tau)S_R(\tau)=p_{R|\beta}^0(\tau){\cal S}(\tau), \\
	&&p_{R|R}(\tau)={\cal K}(\tau)S_R(\tau)=S_R^0(\tau){\cal Q}(\tau).
	\label{WTDRRmemoryresetting}
\end{eqnarray}
In contrast to the case with memoryless resetting, we specifically introduce the time parameter $T$, which denotes the time interval between a collapsed state and the most recent previous reset; see Fig.~(\ref{fig1})(b). Obviously, the breakdown of the sMP is due to the time dependence of the hazard function $K(\tau)$. That is, if the function were constant, these SDs and WTDs would reduce to Eqs.~(\ref{survivaldistributionmemorylessresetting})-(\ref{waitingtimedispRR}).

Although the quantum jump trajectories in the presence of memory resetting are no longer a sMP, the MGF~(\ref{momentgeneratingfunction}) defined by the probabilities of the trajectories is still true. Of course, we need to modify the notation of the trajectories to  
\begin{eqnarray}
	\label{quantumjumptrajectorywithmemory}
	\vec X_N=(	|R\rangle_1,\phi_{\alpha(1,1)},\cdots,|R\rangle_2,\phi_{\alpha(2,1)},\cdots,|R\rangle_N,\phi_{\alpha(N,1)},\cdots),
\end{eqnarray}
where the total number of resets is $N$, $|R\rangle_I$ and $\phi_{\alpha(I,i)}$ denote the reset state and collapsed state at times $t_I$ and $t_{I,i}$, respectively, $I=1,\cdots,N$, and $i=1,\cdots$. On the other hand, we can apply Eqs.~(\ref{survivaldistributionmemorylessresetting})-(\ref{waitingtimedispRR}) to explicitly write the probability density of an arbitrary trajectory, e.g., that in Fig.~(\ref{fig1})(b): 
\begin{eqnarray}
	&&p_{R|\alpha(1,1)}(t_{1,1}-T_1)p_{\alpha(1,1)|R}(T_2-t_{1,1})p_{R|\alpha(2,1)}(t_{2,1}-T_2)p_{\alpha(2,1)|\alpha(2,2)}(t_{2,2,}-t_{2,1},t_{2,1}-T_2)\cdots\nonumber \\
	&&p_{R|\alpha(N,1)}(t_{N,1}-T_N)S_{\alpha(N,1)}(t-t_{N,1},t_{N,1}-T_N).
\end{eqnarray}
However, if we are not simulating quantum jump trajectories, e.g., as in Sec.~(\ref{section5})), these complicated probability formulas are not very useful.

According to probability theory, The probability density of the quantum jump trajectory~(\ref{quantumjumptrajectorywithmemory}) is equal to the product of the probability density ${\mathcal P}[\vec R_N]$ of observing the time series of resets, 
\begin{eqnarray}
	\label{timeseriesofresetting}
	\vec R_N=(|R\rangle_1,|R\rangle_2,\cdots,|R\rangle_N),
\end{eqnarray} 
and the conditional probability density of observing the time series of the collapsed states given the time series of resets. Importantly, if a reset occurred, a segment of the quantum jump trajectory between now and the next reset is independent of the history before this reset. Hence, we formally write the probability density of the quantum jump trajectory~(\ref{quantumjumptrajectorywithmemory}) as 
\begin{eqnarray}
	\label{probabilitydensityofquantumtrajwithmemoryresetting}
	{\mathcal P}[\vec X_N]={\mathcal P}[\vec R_N]\prod_{I=1}^{N}{\mathcal P}_{0}[\vec X_N^I], 
\end{eqnarray}
where ${\mathcal P}_0[\vec X_N^I]$ is the conditional probability density of a segment of the quantum jump trajectory  
\begin{eqnarray}
	\label{quantumjumptrajectorybetweentworesetting}
	\vec X_N^I=(|R\rangle_I,\phi_{\alpha(I,1)},\phi_{\alpha(I,2)},\cdots, \phi_{\alpha(I,M_I)}) 
\end{eqnarray}
given that the $I$-th and $I+1$-th resets happened. Here, we explicitly set the number of collapses in the segment to $M_I$. Obviously, the entire quantum jump trajectory is a combination of these segments; that is, 
\begin{eqnarray}	
	\vec X_N=(\vec X_N^1,\cdots,\vec X_N^N).  
\end{eqnarray}
Using the WTD and SD of the resetting process, we have 
\begin{eqnarray}
	{\mathcal P}[\vec R_N]={\cal Q}(T_2-T_1)\cdots {\cal Q}(T_N-T_{N-1}){\cal S}(t-T_N).
\end{eqnarray}
On the other hand, the conditional probability density ${\mathcal P}_0[\vec X_N^I]$ is simply the probability density of the quantum jump trajectory~(\ref{quantumjumptrajectorybetweentworesetting}) of a reset-free quantum system with the special initial quantum state $|R\rangle$. This case was discussed in Sec.~(\ref{subsection1}). Therefore, we have 
\begin{eqnarray}
	\label{defprobsegmenttraj}
	{\mathcal P}_0[\vec X_N^I]=
	p^0_{R|\alpha(I,1)}(t_{I,1}-T_I)p^0_{\alpha(I,1)|\alpha(I,2)}(t_{I,2}-t_{I,1})\cdots S^0_{\alpha(I,M_I)}(T_{I+1}-t_{I,M_I}). 
\end{eqnarray}
 
Before achieving the desired MGF in the presence of memory resetting, we need to rewrite the time-extensive quantity~(\ref{generalcountingquantity}) as 
\begin{eqnarray}
	\label{sumcountingnumberofsegements}
	C[\vec X_N]=\sum_{I=1}^N  C[\vec X_N^I].
\end{eqnarray}
The reason is simply that we have already set the weights $\omega_{\alpha R}$ and $\omega_{RR}$ to zero. Substituting Eqs.~(\ref{probabilitydensityofquantumtrajwithmemoryresetting}) and~(\ref{sumcountingnumberofsegements}) into the definition of the MGF, Eq.~(\ref{momentgeneratingfunction}), we have  
\begin{eqnarray}
	\label{MGFdefinitionmemoryresetting}
 	M(\lambda,t)
 	&=&\sum_{N=1}^\infty\sum_{\vec X_N} {\cal P}[\vec R_N]\prod_{I=1}^N e^{-\lambda C[\vec  X_N^I]} {\cal P}_0[\vec X_N^I] \nonumber\\
 	&=&\sum_{N=1}^\infty\sum_{\vec R_N}{\cal P}[\vec R_N]\sum_{\vec X_N^1}\cdots\sum_{\vec X_N^N}\prod_{I=1}^N e^{-\lambda C[\vec X_N^I]} {\cal P}_0[\vec X_N^I]\nonumber\\
 	&=&\sum_{N=1}^\infty\sum_{\vec R_N}{\cal P}[\vec R_N]\prod_{I=1}^N\left(\sum_{\vec X_N^I} e^{-\lambda C[\vec X_N^I]} {\cal P}_0[\vec X_N^I]\right). 
\end{eqnarray} 
We see that 
the term in parentheses in the last equality 
is simply the MGF $M_{0R}(\lambda,T_{I+1}-T_I)$ discussed in Sec.~(\ref{subsection1}). Note that the summation over the time series of resetting, Eq.~(\ref{timeseriesofresetting}), is in fact a shorthand notation for the time-ordered integrals at different times: 
\begin{eqnarray}
	\label{timeintegrals}
	\sum_{\vec R_N}\equiv\int_0^tdT_2\int_{T_2}^{t}dT_3\cdots\int_{T_{N-1}}^{t}dT_N.
\end{eqnarray}
The complex Eq.~(\ref{MGFdefinitionmemoryresetting}) can be dramatically simplified if we take its Laplace transform, and we arrive at 
\begin{eqnarray}
	\label{MGFmemoryresetting}
	\hat M(\lambda,v)=\frac{(\hat {\cal S}*\hat M_{0R})(\lambda,v)}{2\pi {\rm i}-(\hat {\cal Q} *\hat M_{0R})(\lambda,v)},
\end{eqnarray}
where the asterisks represent convolutions and Eq.~(\ref{defprobsegmenttraj}) is used.

Equation~(\ref{MGFmemoryresetting}) has two consequences. First, we can obtain the SCFG in the presence of memory resetting by finding the largest real root of an algebraic equation with the parameter $v$: 
\begin{eqnarray}
\label{SCGFformulaformemoryresetting}
\frac{1}{2\pi {\rm i}}(\hat{\cal  Q} *\hat M_{0R})(\lambda,v)-1=0.
\end{eqnarray}
Second, Eq.~(\ref{MGFmemoryresetting}) can be interpreted as a result of a matrix equation analogous to Eq.~(\ref{tiltedmatrixequationwithmemorylessrestting}), but the tilted matrix therein is updated to  
\begin{eqnarray}
	\label{tiltedmatrixmemoryresetting}
	{\bf G}(v)=\left[\left(\hat{ {\cal S}}* {\bf G}_{0R}^{-1}\right)(v)\right]^{-1}\left[{2\pi{\rm i} }- \left({\cal  \hat Q} *  {\bf G}_{0R}^{-1}\right)(v){\bf \Pi}\right].
\end{eqnarray}
We may simply verify Eqs.~(\ref{MGFmemoryresetting})-(\ref{tiltedmatrixmemoryresetting}) by applying them to the memoryless resetting case: because the WTD and SD of such a resetting process are exponential decay functions with the constant $K$ and the convolutions therein are proportional to $\hat M_{0R}(v+K)$ and ${\bf G}_{0R}^{-1}(v+K)$, Eqs.~(\ref{tiltedmatrixmemorylessresetting}),  (\ref{MGFmemorylessrestting}), and (\ref{SCFGmemorylessresetting}) can be rederived. 

We close this section by commenting on differences between our theory and that of Perfetto et al.~\cite{Perfetto2022}. 
They obtained Eq.~(\ref{MGFmemoryresetting}) by a method that is entirely based on the TQME~\cite{Mollow1975,Esposito2009,Garrahan2010}. Although our tilted matrix equation~(\ref{MEofcountingstatisticsmatrixform}) was proven to be equivalent to the TQME, this consistency only holds for a special type of time-extensive quantity, i.e., the weights that depend only on the second collapse of a pair of collapsed states~\cite{Liu2022}. From this perspective, our results are more general than the previous ones. On the other hand, the mathematics we are using is essentially classical probability theory. In contrast, Perfetto et al.~\cite{Perfetto2022} applied a hierarchy of equations about conditional density matrices. The probability meanings in their method are not very direct. In the next section, we will show that the classical probability formulas are very useful when we attempt to simulate SCGFs of open quantum systems, either with or without resetting. 
  
\section{ Continuous-time cloning algorithm} 
\label{section5}
The algebraic Eq.~(\ref{SCFGmemorylessresetting}) or the more general algebraic Eq.~(\ref{SCGFformulaformemoryresetting}) provides us with a way of calculating SCGFs of open quantum systems with resetting. The first step is to solve the reset-free MGF $\hat M_{0R}(\lambda,v)$ with the special initial state $|R\rangle$. The next step is to solve the algebraic equations. In general, these two steps are implemented numerically. On the other hand, Cavallaro and Harris~\cite{Cavallaro2016} developed a continuous-time cloning algorithm (CTCA) to simulate SCGFs of an arbitrary classical non-Markov process. In this paper, we do not review the CTCA of Cavallaro and Harris. Interested readers are referred to the original article~\cite{Cavallaro2016}.  Because we have established the sMP theoretical framework for quantum jump trajectories either with or without resetting, which is a special case of non-Markov processes, we can directly apply their algorithm to the current situation. The key ingredients of the simulation are presented in Eqs.~(\ref{SDalphamemoryresetting})-(\ref{WTDRRmemoryresetting}). 

\section{Several examples } 
\label{section6}
\subsection{Reset-free resonant two-level system}
\begin{figure}
\includegraphics[width=1\columnwidth]{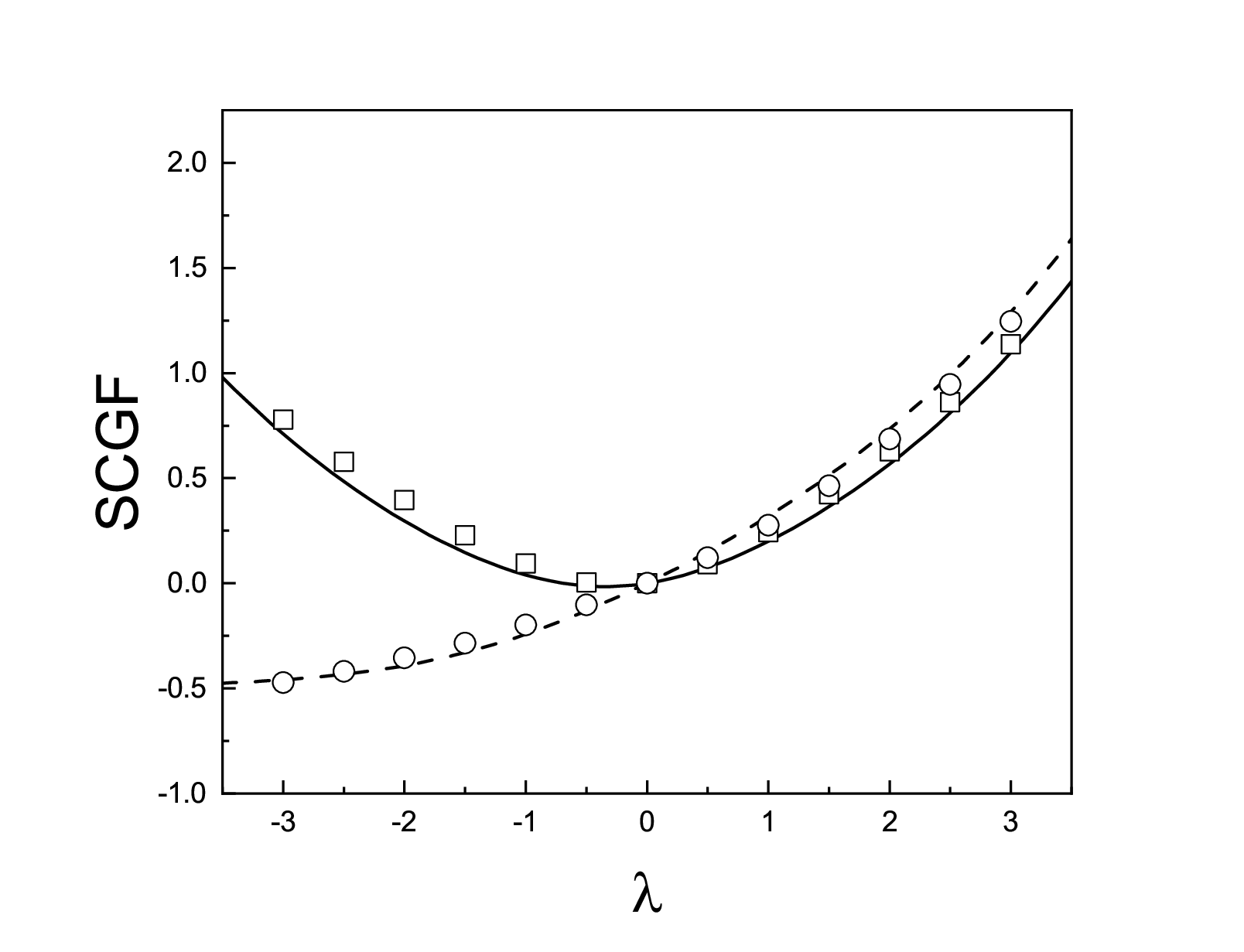}
	\caption{The solid and dashed curves are the SCGFs simulated by the CTCA. The open squares and circles are the exact numeric data calculated by Eq.~(79) in Ref.~\cite{Liu2022}. Two sets of parameters are used: for the solid curve and squares, $r_-=1$, $r_+=0.5$, $\Omega=0.8$, and $\omega_0=1$; for the dashed curve and circles, $r_-=1$, $r_+=0.0$, $\Omega=0.8$, and $\omega_0=1$. The latter concerns a TLS in a vacuum. These parameters were also applied in a previous paper; see Fig.~(2) therein~\cite{Liu2022}. In the simulations, the number of clones is 2000, and the simulation time is 1500. }
	\label{fig2}
\end{figure}

We first give an example of the CTCA by simulating SCGFs in a reset-free open quantum system. To the best of our knowledge, applications of the algorithm in open quantum systems are rare in the literature. We choose a resonant two-level system (TLS) whose SCGF has an exact expression~\cite{Liu2022}. The quantum system is driven by a resonant field and surrounded by an environment with inverse temperature $\beta$. In the interaction picture, the MQME of the TLS~\cite{Mandel1995} is
\begin{eqnarray}
	\label{QMEtwolevel}
	\partial_t\rho(t)&=&-{\rm i}\left[ H,\rho(t)\right ] + r_-[\sigma_-\rho(t)\sigma_+ -\frac{1}{2}\{\sigma_+\sigma_-, \rho(t) \}   ]\nonumber \\
	&&+r_+[\sigma_+\rho(t)\sigma_- -\frac{1}{2}\{\sigma_-\sigma_+, \rho(t) \}   ].
\end{eqnarray}
Here, $H=-\Omega(\sigma_- +\sigma_+)/2$ represents the interaction Hamiltonian between the system and the resonant field, $\sigma_\pm$ are the raising and lowering Pauli operators, $\Omega$ is the Rabi frequency, and $r_\pm$ are the pumping and damping rates. The two rates satisfy the detailed balance condition, $r_-=r_+\exp{(\beta\omega_0 )}$, and $\omega_0$ is the energy level difference of the two-level system. There are two collapsed states: the ground state $|0\rangle$ and the excited state $|1\rangle$. We select heat production as a time-extensive quantity, the weights of which are 
\begin{eqnarray}
\label{weightsofheat}
	\{\omega_{00},\omega_{01},\omega_{10},\omega_{11}\}\rightarrow\{\omega_0,-\omega_0,\omega_0,-\omega_0\}.
\end{eqnarray}
Because the WTDs and SDs of the simple system are exactly known~\cite{Liu2022}, the CTCA is easily implemented. Figure~(\ref{fig2}) shows the simulated SCGFs under two sets of parameters. For comparison, exact numeric data are shown in the same figure. We see that their agreement is very satisfactory.  

\subsection{Memoryless resetting }
In the presence of resetting, we focus on a TLS in a vacuum; i.e., $r_+=0$ in Eq.~(\ref{QMEtwolevel}). A similar system was considered by Perfetto et al.~\cite{Perfetto2021}. Unlike the previous discussions, which fully rely on numerical schemes, we remain as analytically accessible as possible. Here, there is only one collapsed state, the ground state $|0\rangle $. Let the reset state be  
\begin{eqnarray}
	\label{resettingstateofTLS}
	|R\rangle=a|0\rangle + b|1\rangle.
\end{eqnarray} 
For simplicity, the parameters $a$ and $b$ are assumed to be real numbers. We select the time-extensive quantity~(\ref{generalcountingquantity}) as the number of collapses to the ground state; that is, the unique weight $\omega_{00}=1$. Performing direct calculations, we obtain the Laplace transform of the reset-free MGF with the initial state~(\ref{resettingstateofTLS}):  
\begin{eqnarray}
	\label{MGFTLS-resettinginitialstate-resettingfree}
\hat M_{0R}(\lambda,v)&=&\hat{P}_R(v)+	\hat{P}_0(v)\nonumber \\
&=&\hat{S_R^0}(v) + \frac{\hat{S_0^0}(v)}{1-e^{-\lambda}\hat p_{0|0}^0(v)}\hat p_{R|0}^0(v) e^{\lambda}
\end{eqnarray} 
where the reset-free WTDs and SDs with the special initial states $|R\rangle$ and $|0\rangle$ are connected by  
\begin{eqnarray}
	\hat S_R^0(v)&=&\hat S_0^0(v)-b^2\frac{r_-}{\xi^2+4\mu^2}\nonumber\\
	&=&\frac{\xi^2+r_-(\xi+r_-/2)/2+4\mu^2}{\xi(\xi^2+4\mu^2) }-b^2\frac{r_-}{\xi^2+4\mu^2}, \label{SDresettingfreeRstateandgroundstate} \\
	\hat{p}^0_{R|0}(v)&=&\hat{p}^0_{0|0}(v)+b^2\frac{r_-(\xi-r_-/2)}{\xi^2+4\mu^2}, \nonumber\\
	&=&\frac{r_-\Omega^2}{2\xi(\xi^2+4\mu^2)}+b^2\frac{r_-(\xi-r_-/2)}{\xi^2+4\mu^2}, 
	\label{WTDresettingfreeRstateandgroundstate}
\end{eqnarray}
respectively, where $\xi\equiv v+r_-/2$ and  $16\mu^2\equiv 4\Omega^2-r_-^2$ ($>0$). If $b^2$ is equal to zero, Eq.~(\ref{MGFTLS-resettinginitialstate-resettingfree}) is the Laplace transform of the reset-free MGF $\hat M_0(\lambda,v)$. 
  
\begin{figure}
\includegraphics[width=1\columnwidth]{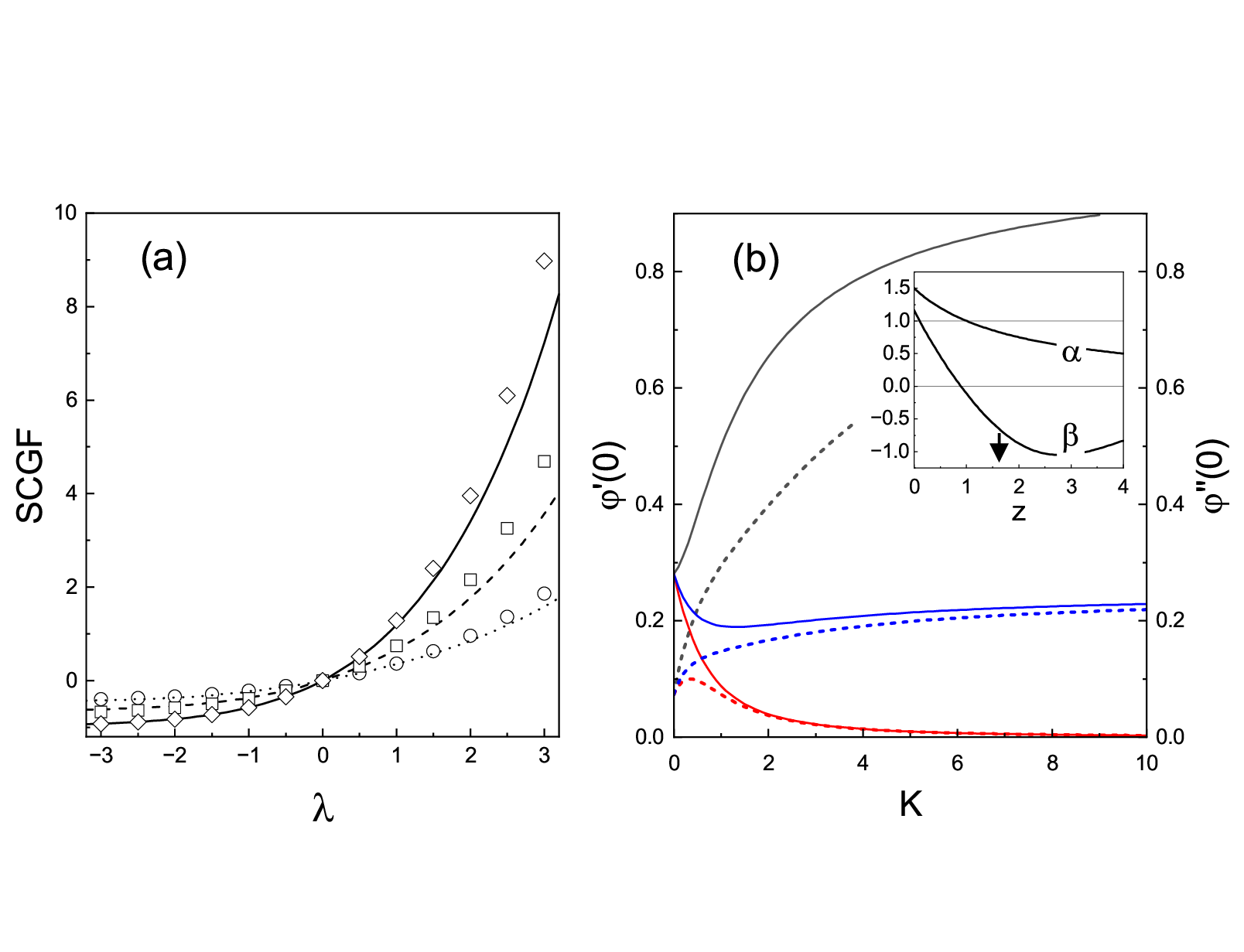}
	\caption{(a) The solid, dashed, and dotted curves are the SCGF data solved by Cardano's formula. The symbols are the simulated data given by the CTCA. The parameters are $r_-=1$, $\Omega=0.8$, and $\omega_0=1$, and the rates $K$ are $5$, $1$, and $0.1$, accordingly. The reset state is fixed at the excited state $|1\rangle$; i.e., $b=1$. The number of clones is 2000, and the simulation time is 1500. (b) The solid and dashed curves are the first and second derivatives of the SCFG at $\lambda=0$, respectively. The parameters $b^2$ in the reset states are $1$, $1/2$, and $0$ for the dark, blue, and red curves, respectively. Inset: The curves of $\alpha$ and $\beta$; see Eqs.~(\ref{alphafunc}) and~(\ref{betafunc}). The arrow indicates the $z$-value of the parameters.}
	\label{fig3}
\end{figure}

To calculate the SCGF of large deviations of the open quantum system with memoryless resetting, we substitute Eq.~(\ref{MGFTLS-resettinginitialstate-resettingfree}) into Eq.~(\ref{SCFGmemorylessresetting}) and simplify to obtain an algebraic equation involving $v$:    
\begin{eqnarray}
	\label{cubicequationmemorylessresetting}
	\zeta^3(v)-K\zeta^2(v)+\left[4\mu^2-
	\frac{1}{2}Kr_-+b^2Kr_-\left(1-e^{-\lambda}\right)\right]\zeta(v)-\Omega^2\left(K+\frac{1}{2}r_-e^{-\lambda}\right)=0, 
\end{eqnarray} 
where $\zeta(v)=v+r_-/2+K$. If the rate $K$ is zero, Eq.~(\ref{cubicequationmemorylessresetting}) reduces the cubic equation for the SCGF of the reset-free TLS in a vacuum [Eq. (C8) in Ref.~(\cite{Liu2022})] and does not involve the parameter $b$. It is expected that the zero rate implies that resetting is absent from the dynamics of the quantum system, and the long-time behavior of the system is independent of the initial quantum state. The SCGF is solved by finding the largest real root of Eq.~(\ref{cubicequationmemorylessresetting}). This is a cubic algebraic equation and has an exact solution given by Cardano's formula~\cite{Speigel1968}. Considering that the formula is slightly lengthy, in Fig.~(\ref{fig3})(a), we only show its exact numerical values under several sets of parameters and compare them with the data simulated by the CTCA. We see that the two methods are indeed consistent with each other.  

Equation~(\ref{cubicequationmemorylessresetting}) provides us with intriguing information about the statistics of the counting current $j$. For instance, through it, we can derive analytical expressions for the first and second derivatives of the SCGF at $\lambda=0$:
\begin{eqnarray}
	\varphi'(0)&=&\zeta'(0)=\frac{r_-[\Omega^2+2b^2K\zeta(0)]}{8\mu^2-Kr_--4K\zeta(0)+6\zeta(0)^2}, \label{meancurrent}\\
	\varphi''(0)&=&\frac{\Omega^2r_-+4(K-3\zeta(0))\zeta'(0)^2+2b^2Kr_-[\zeta(0)+2\zeta'(0)]}{8\mu^2+6\zeta(0)^2-K(r_-+4\zeta(0))},
	\label{variancecurrent}
\end{eqnarray}
where $\zeta(0)=K+r_-/2$. According to the large deviation theory~\cite{Touchette2008}, Eq.~(\ref{meancurrent}) is the mean current, while Eq.~(\ref{variancecurrent}) indicates the fluctuation of the current in the long time limit (the coefficient of diffusivity). In Fig.~(\ref{fig3})(b), we show their values at different resetting rates with different reset states. We see that at larger $K$ values, all of them tend toward certain values. At smaller and intermediate $K$ values, however, their behaviors are diverse and depend on the concrete values of the parameters.

It is not trivial to present concise and clear explanations for the various behaviors of the mean and fluctuation of the counting current under general parameters, e.g., the nonmonotonic phenomena of the fluctuation with $b^2=0$ and mean current with $b^2=1/2$ in Fig.~(\ref{fig3})(b). These complexities arise from mutual matching and/or competition among many factors, including the resetting frequency, different WTDs with respect to different reset states or collapsed states, etc. Hence, the following discussion is restricted to several of the simplest cases. We know that in the quantum jump trajectories, the initial state of the deterministic quantum processes is either a reset state or a collapsed state (here, only the ground state). It takes a certain amount of time to evolve from these quantum states to collapse~\cite{Breuer2002}. We can prove that the higher the probability ($b^2$) of the excited state in these quantum states is, the shorter the time~\footnote{The mean waiting time is $\bar{\tau}_{R/0}=\hat S_{R/0}(0)$. Equation~(\ref{SDresettingfreeRstateandgroundstate}) immediately leads to this conclusion.}. When the probability is negligible in the reset state, resetting only decreases the rate of collapse. This is because resetting interrupts the deterministic processes from the collapsed state to the next collapse. Furthermore, the time required for the deterministic process from this reset state to the next collapse is almost the same as the time required for the previous quantum process. Therefore, the more frequently the quantum system resets, the smaller the rate at which the system collapses.
In contrast, if the probability of the excited state is dominant in the reset state, resetting will increase the rate of collapse. Although resetting indeed interrupts the deterministic processes from the collapsed state to the next collapse, the time required for the deterministic process from the reset state to the next collapse is shorter than that of the previous quantum process. Overall, it still shortens the time interval between the two collapses. Therefore, the more frequently the quantum system resets, the higher the rate at which the system collapses. Finally, if the resetting rate is so frequent that resetting even interrupts the deterministic processes from the previous reset state to the next collapse, it seems that the quantum system is almost ``frozen" at the reset state. The probability of a collapse from the reset state during a small time interval $\Delta t$ is proven to be $b^2r_-\Delta t$~\footnote{According to the Laplace initial value theorem, the WTD $p_{R|0}(t)$ at time $0$ is $p_{R|0}(0)=\lim_{v\rightarrow\infty}v\hat p_{R|0}^0(v)=b^2r_-$; see Eq.~(\ref{WTDresettingfreeRstateandgroundstate}) }. In this situation, the quantum jump trajectories are close to the Poisson counting process with a constant rate $b^2r_-$. Accordingly, the fluctuation is equal to $b^2r_-$. These discussions qualitatively explain the asymptotic behaviors of all curves at very large $K$ values and the mean current curves with $b^2=0,1$ at smaller $K$ values in Fig.~(\ref{fig2})(b). 

The previous discussions can be treated in a quantitative way. The asymptotic behaviors are easily seen from Eqs.~(\ref{meancurrent}) and~(\ref{variancecurrent}): when $K\rightarrow\infty$, both $\varphi'(0)$ and $\varphi''(0)$ tend toward $b^2r_-$. At smaller $K$ values, we use the Taylor expansion to analyze the variation: 
\begin{eqnarray}
	\label{meancurrentlinearapprox}
	\varphi'(0)&\approx& \frac{\alpha }{3}r_-+\frac{z}{2+z}(b^2-\alpha)K, \\
 	\varphi''(0)&\approx& \frac{(2-z)^2+2 z}{(2 +z)^3}r_-+\frac{z[(2 -z)^2 +8]}{(2+z)^3}\left(b^2-\beta\right)K, 
 \label{fluctcurrentlinearapprox}
\end{eqnarray}
where the dimensionless $z$ is equal to $r_-^2/\Omega^2$ ($<4$) and 
\begin{eqnarray}
\alpha&=&\frac{3}{2 +z},
\label{alphafunc}\\
\beta&=&\frac{28 -34z+3z^2}{(2+z)[(2 -z)^2 +8 ]}. 
\label{betafunc}
\end{eqnarray}
The first terms on the right-hand side of Eqs.~(\ref{meancurrentlinearapprox}) and~(\ref{fluctcurrentlinearapprox}) are the mean and fluctuation of the counting current without resetting, respectively. The dependence of $\alpha$ and $\beta$ on $z$ are plotted in the inset of Fig.~(\ref{fig2})(b). For the parameters applied in the figure (see the bold arrow therein), $0<\alpha<1$, $\beta<0$, and the curves of fluctuation increase at $K=0$, while for the curves of the mean current, there is a change of the slopes at $K=0$ from negative to positive. In addition, Eq.~(\ref{meancurrentlinearapprox}) implies nonmonotonic features of the mean current when $b^2$ is within the interval $(\alpha/3,\alpha)$, while Eq.~(\ref{fluctcurrentlinearapprox}) also implies that the fluctuation curve with $b^2=0$ must have a maximum value.

\subsection{Memory resetting}
To illustrate memory effects in the counting statistics with resetting, we choose the Erlang-$2$ distribution with rate parameter $K$ as the WTD of the resetting process: 
\begin{eqnarray}
	Q(\tau)=K^2 \tau e^{-K \tau}.
\end{eqnarray} 
Note that the mean rate of this distribution (the reciprocal of the mean waiting time) is equal to $K/2$, and the variance is equal to $2/K^2$. Substituting this distribution into Eq.~(\ref{SCGFformulaformemoryresetting}), we have
\begin{eqnarray}
	\label{ErlangSCGFequation}
K^2 \frac{d}{d\zeta} \hat M_{0R}(\lambda,\zeta)+1=0,
\end{eqnarray}
$\zeta(v)=v+r_-/2+K $, and the reset-free MGF with the initial quantum state $|R\rangle$ is given by Eq.~(\ref{MGFTLS-resettinginitialstate-resettingfree}). Although the equation is exact, a simplification shows that this is a 
sixth-order
 algebraic equation about $v$, and a numerical scheme must be used to find the largest real roots of $v$ given $\lambda
$. We present the data for several $K$-values in Fig.~(\ref{fig3})(a) and compare them with the SCGF data simulated by the CTCA. We see that their agreements are also satisfactory. 

\begin{figure}
\includegraphics[width=1\columnwidth]{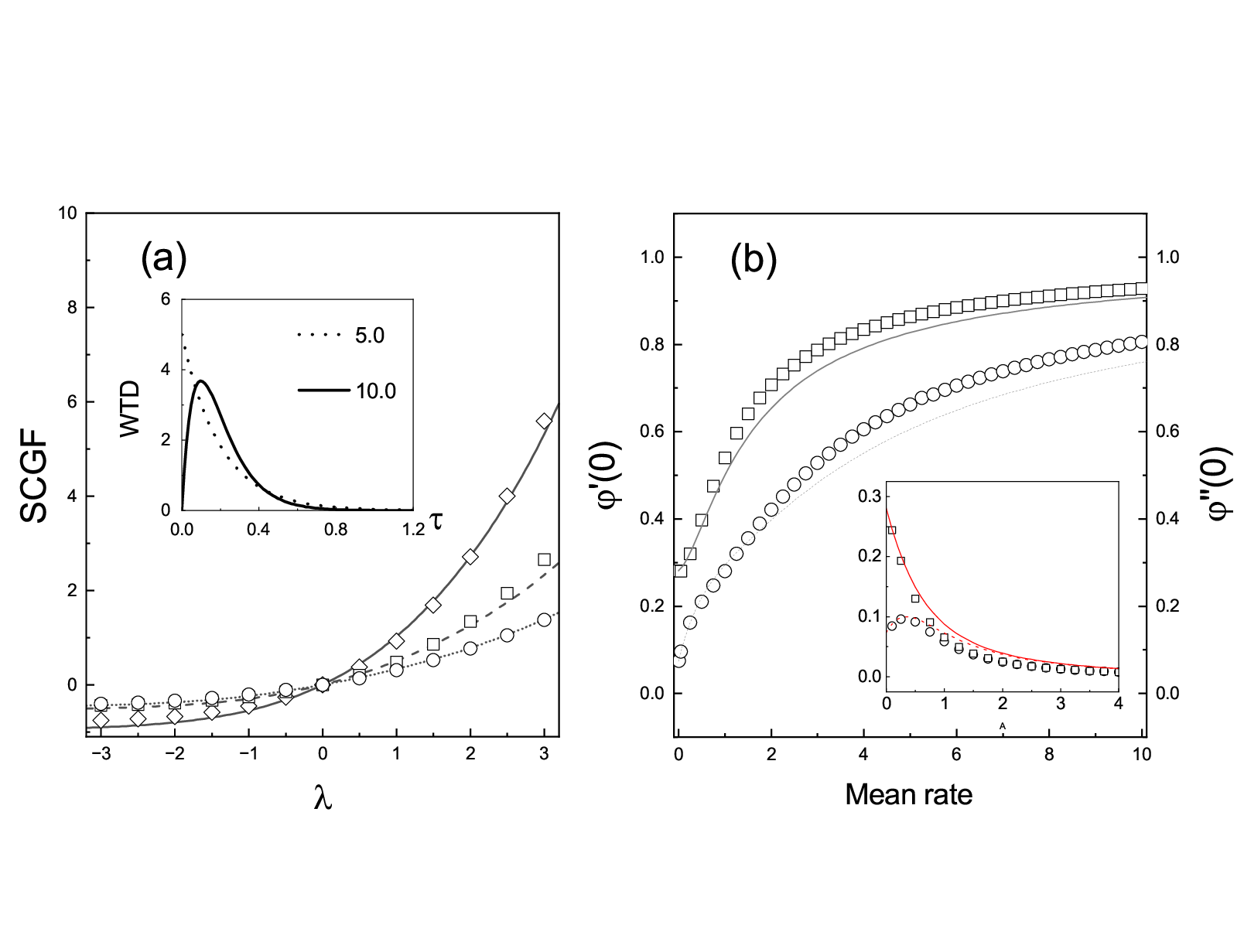}
	\caption{(a) The solid, dashed, and dotted curves are the SCGF data obtained by solving Eq.~(\ref{ErlangSCGFequation}) by a numerical method. The symbols are the data simulated by the CTCA. The parameters are $r_-=1$, $\Omega=0.8$, and $\omega_0=1$, and the rates $K$ are $5$, $1$, and $0.1$, accordingly. Here, the reset state is set to the excited state. The number of clones is 2000, and the simulation time is 1500. Inset: A comparison between an exponential distribution with rate $5$ (dashed curve) and an Erlang distribution with rate parameter $K=10$ (solid curve). Note that their mean rates are the same, and the variance of the latter ($2/100$) is smaller than that of the former ($1/25$). (b) The squares and circles are the first and second derivatives of the SCFG at $\lambda=0$, respectively. Here, the reset state is the excited state. For comparison, we also replot the black solid and dashed curves in Fig.~(\ref{fig3})(b) for the case of memoryless resetting. Note that we show the mean rate on the horizontal axis. Inset: Analogous data where the reset state is the ground state.}
	\label{fig4}
\end{figure}

Similar to the memoryless-resetting case, Eq.~(\ref{ErlangSCGFequation}) provides useful information about memory effects on the mean and fluctuation of the counting current. Rather than writing very lengthy equations analogous to Eqs.~(\ref{meancurrent}) and~(\ref{variancecurrent}), we directly present their numerical values in Fig.~(\ref{fig4})(b) and the inset with special reset states. To compare the data of the memoryless-resetting case, we use the mean rate as the horizontal axis. We find that these mean curves and fluctuation curves with memory are similar to the previous ones without memory. At smaller and adequately large mean rates, they almost overlap. Their differences become apparent at intermediate $K$ values: the mean current and fluctuation with memory are larger than those without memory if the reset state is the excited state, while the opposite conclusion is obtained if the reset state is the ground state; see the inset. This demonstrates the complexity of the interactions between memory and the reset state in quantum jump trajectories.

At adequately small and large $K$ values, memory is marginal in the counting statistics. The latter point is easily seen since the quantum state is almost ``frozen" in the reset state, as resetting is very frequent. For smaller $K$, we apply Taylor expansion again and find that the mean current has the same expression as Eq.~(\ref{meancurrentlinearapprox}) except that the parameter $K$ therein is replaced by $K/2$. Because it is simply the mean rate of the Erlang distribution, we explain the consistency of the mean currents with memoryless and memory resetting. Regarding the fluctuation, we obtain a slightly complicated expression: 
\begin{eqnarray}
	\label{Erlangmeancurrentlinearapprox}
 \varphi''(0)&\approx& \frac{(2-z)^2+2 z}{(2 +z)^3}r_-+\\ \nonumber 
 && \frac{z[(2 -z)^2 +8+3z]}{ (2+z)^3}\left[b^2-\frac{28-(59/2)z+3z^2}{(2+z)[(2-z)^2+8+3z ]}-b^4\frac{z(z+2)}{2[(2-z)^2+8+3z]}\right]\frac{K}{2}. 
\label{Erlangfluctcurrentlinearapprox}
\end{eqnarray}
The $b^4$-term clearly indicates memory effects on the fluctuations. Although Eqs.~(\ref{fluctcurrentlinearapprox}) and~(\ref{Erlangfluctcurrentlinearapprox}) are distinct, the calculations show that their data are close, especially for the parameters applied in the figure (given the same mean rate, the values with memory are slightly smaller than the values without memory).



\section{Conclusion}  
\label{section7}
In this paper, we extend our previous sMP method for determining the counting statistics of open quantum systems to situations with memoryless and memory resetting. For the former situation, because the composition of the random events, which includes the collapses of the wave function and the quantum reset state, is still a sMP, the method can be directly applied by simply adding the reset state into the set of collapsed states. For the latter situation, the composite stochastic process is no longer a sMP. Even so, because resetting affects quantum processes only through the initial quantum states instead of altering the quantum dynamics, using probability formulas, we prove that the MGF of open quantum systems with memory resetting can be calculated by relating it to the MGF of reset-free open quantum systems. Although this conclusion agrees with the previous one, its validity has been expanded to general counting statistics. A tilted matrix equation that has not been previously discovered is proposed. Finally, to simulate the large-deviation statistics of general random time-extensive quantities, on the basis of a set of probability formulas that can characterize the composite stochastic process, we introduce the CTCA to open quantum systems. To illustrate these theoretical results, we concretely calculate the large-deviation properties and the SCGFs of two-level quantum systems. On the one hand, we verify that the CTCA is quite accurate by comparing the simulation data to the exact analytical or numerical results. On the other hand, we also find that the effects of resetting on quantum systems can be very complex, even if resetting is memoryless. The plausible reason is that the waiting-time distributions of reset-free systems are not trivial at all; e.g., there are quantum antibunching effects. The presence of resetting, especially memory resetting, further increases this complexity. Hence, quantitative formulas are more trustworthy than qualitative arguments. For the relatively simple TLS and Erlang-2 distribution, the sMP method is applicable. It will be interesting to investigate the applications of the sMP method in complex quantum many-body systems, in both reset-free and reset cases.

{\noindent\it Acknowledgments}
We thank Dr. Cavallaro for his inspiring discussions of the continuous-time cloning algorithm of non-Markov jump processes. This work was supported by the National Natural Science Foundation of China under Grant Nos. 12075016 and 11575016.\\

\appendix

\section{Renewal equation of the density matrix}
Perfetto et al.~\cite{Perfetto2022} derived a renewal equation relating the dynamics of the reduced density matrix $\rho(t)$ in the presence of memory resetting to the reset-free matrix and called the evolution equation the generalized Lindblad quantum master equation. It is interesting to connect the renewal equation with the more fundamental probability formulas and wave function notion developed in this paper. First, the reduced-density matrix of the quantum system at time $t$ can be expressed by the age structure~\cite{Liu2022}:  
\begin{eqnarray}
	\label{densitymatrixageform}
	\rho(t)=\sum_{\alpha=1}^{M}\int_0^t d\tau p_\alpha(t,\tau) U(\tau)|\phi_\alpha\rangle\langle \phi_\alpha|U^\dag(\tau)
	+\int_0^t d\tau p_R(t,\tau) U(\tau)|R\rangle\langle R|U^\dag(\tau), 
\end{eqnarray}
where $U(\tau)$ is the time-evolution operator of the nonlinear Schr$\ddot{o}$dinger equation~\cite{Breuer2002}. In Eq.~(\ref{densitymatrixageform}), $p_\alpha(t,\tau)$ represents the probability density that the quantum system starts from the collapsed state $\phi_\alpha$ at time $t-\tau$ and continuously evolves until time $t$. Thus, the age of the system is $\tau$. The meaning of the probability density $p_R(t,\tau)$ is similar except that the system starts from the reset state $|R\rangle$. It is easier to understand their meanings by referring to Fig.~(\ref{fig1}).  

Because of memory effects, the possibility of the continuous evolution of a quantum state is affected by the time interval $s$ from the last reset to the current moment~\footnote{We introduced the parameter $T$ in Sec.~\ref{section4}. These time parameters are related by $T=s-\tau$. }. Hence, it is useful to write $p_\alpha(t,\tau)$ in a detailed way: 
\begin{eqnarray}
	p_\alpha(t,\tau) =\int_\tau^t  p_\alpha(t,\tau,s)ds.
\end{eqnarray}
Following the idea of Eq.~(\ref{probabilitydensityofquantumtrajwithmemoryresetting}), we may temporally overlook all collapses and focus only on the resetting process. Let the probability density $P(t,s)$ represent that there have been no resets at time $t$ since the last reset at time $t-s$. That is, the time without resets is $s$. Then, according to the probability theory, we can rewrite  
\begin{eqnarray}
	\label{A3}
	p_\alpha(t,\tau,s)=P(t,s)p^{0}_{R\alpha}(s,\tau),  \end{eqnarray}
where $p^{0}_{R\alpha}(s,\tau)$ is the conditional probability density that the quantum system continuously evolves until time $s$ with age $\tau$ ($<s$). Note that the subscript $R$ on the right-hand side denotes that the reset state $|R\rangle$ is the initial quantum state at $s=0$. Analogously, the other probability density in Eq.~(\ref{densitymatrixageform}) is rewritten as 
\begin{eqnarray}
	\label{A4}
	p_R(t,\tau)&=&P(t,\tau)S^{0}_R(\tau).
	\end{eqnarray}
Substituting Eqs.~(\ref{A3}) and~(\ref{A4}) into Eq.~(\ref{densitymatrixageform}) and rearranging the integrals, we obtain
\begin{eqnarray}
	\label{renewalequationdensitymatrix}
	\rho(t)=\int_0^t P(t,s) \rho_{0}(s), 
\end{eqnarray}
where
\begin{eqnarray}
	\label{densitymatrixresettingfree}
	\rho_{0}(s)=\sum_{\alpha=1}^{M}\int_0^s d\tau p_{R\alpha}^0(s,\tau) U(\tau)|\phi_\alpha\rangle\langle \phi_\alpha|U^\dagger (\tau) +S^0_R(s)U(s)|R\rangle\langle R|U^\dagger(s).
\end{eqnarray}
Equation~(\ref{renewalequationdensitymatrix}) is simply the renewal equation, and Eq.~(\ref{densitymatrixresettingfree}) is the reduced-density matrix solution to the MQME~(\ref{MQME}) with the special initial density matrix $|R\rangle\langle R|$~\cite{Liu2022}, which is formally equal to    
\begin{eqnarray}
	\rho_{0}(s)\equiv e^{s\cal L}\left[|R\rangle\langle R|\right].
\end{eqnarray}
In fact, if the probability density $P(t,s)$ is defined from the beginning, the desired equation can be intuitively written out, as Perfetto et al. did previously~\cite{Perfetto2022}. Using the age structure of the resetting process, we can further derive the generalized Lindblad quantum master equation. Considering that this procedure is the same as the previous ones~\cite{Perfetto2022,Eule2016}, we do not show it in this paper. 

\end{document}